\documentclass[reprint,twocolumn,notitlepage,amsmath,amssymb,aps]{revtex4-1}

\usepackage{graphicx}
\usepackage{dcolumn}
\usepackage{bm}

\begin{document}

\title{Renormalization-Group Fixed Points Not Associated with Universality Classes}

\author{Ko Okumura}
 \altaffiliation{Department of Physics and Soft Matter Center, Ochanomizu University, 2-1-1, Ohtsuka, Bunkyo-ku, Tokyo 112-8610, Japan}

\date{\today}

\begin{abstract}

As in critical phenomena, universality classes have recently been identified
for self-similar solutions of nonlinear partial differential equations
(PDEs) as consequences of repeated renormalization-group (RG)
transformations that eliminate irrelevant structures. Whether all RG fixed
points are associated with universality classes, however, remains unclear.
We address this question by extending a recently proposed 'predictive RG
framework' that works for deterministic PDEs to traveling-wave solutions,
using the Burgers and Korteweg--de Vries (KdV) equations as representative
examples.

It has been known that traveling waves are mapped onto self-similar forms by
means of a logarithmic transformation. Accordingly, applying the predictive
RG to the log-transformed governing equations yields an RG-flow analysis
that reproduces the classical shock-wave and soliton solutions, thereby
identifying traveling waves as RG fixed points and providing an
interpretation of the logarithmic variable as RG-flow time. 

Although the predictive RG framework has shown that self-similar fixed
points are associated with universality classes, we find that the present
traveling-wave fixed points do not produce the irrelevant structures whose
elimination normally gives rise to universality classes. This failure of
universality-class formation originates from a distinctive property of the
logarithmic representation: the field and all of its derivatives acquire the
same scaling dimension. The fixed-point existence condition requires this
common scaling dimension to vanish ($A=0$), rendering any analytic local
function of the field and its derivatives scale invariant and preventing the
emergence of irrelevant structures. Traveling waves thus provide explicit
examples of RG fixed points without associated universality classes. These
observations further suggest that the absence of universality classes may be
a generic feature of traveling-wave RG fixed points formulated through
logarithmic coordinates, rather than a peculiarity of the Burgers and KdV
equations.

The absence of universality classes suggests that information associated
with initial conditions may persist asymptotically, thereby promoting a
tendency toward memory retention. Correspondingly, the Burgers and KdV
examples suggest that memory of initial conditions can survive RG
transformations, although the retained information remains system dependent.

\end{abstract}

\maketitle

\section{Introduction}

Universality classes are among the most fundamental concepts in statistical
physics and critical phenomena. A recently developed predictive
renormalization-group (RG) framework has identified universality classes for
self-similar solutions of nonlinear partial differential equations (PDEs) 
\cite{Okumura2025RG, okumura2026combined, Okumura2026nonlinear}. The
framework places self-similar asymptotic solutions within the same
conceptual structure that underlies universality and critical phenomena \cite%
{Cardy, Goldenfeld}, identifying them as RG fixed points and universality
classes as consequences of irrelevant-structure elimination. In doing so, it
brings together three traditions that were originally developed largely
independently: field-theoretic renormalization methods for nonlinear
diffusion problems \cite{goldenfeld1989intermediate, Goldenfeld}, RG
approaches based on repeated rescaling transformations \cite%
{bricmont1994renormalization}, and dynamical-system analyses of self-similar
solutions \cite{giga1985asymptotically, eggers2015singularities}. Recent
work within the same framework has also shown how memory of initial
conditions and system parameters may survive repeated RG transformations 
\cite{Okumura2026memory}.

Whether all RG fixed points are associated with universality classes,
however, remains unclear. Traveling waves provide a natural setting in which
to address this question. They arise in a wide variety of nonlinear
phenomena, including chemical reactions, biological invasions, shock
formation, and dispersive wave propagation, and constitute one of the most
important asymptotic structures of nonlinear partial differential equations
(PDEs) \cite{murray2002mathematical1, murray2002mathematical2,
eggers2015singularities}. Their close connection to self-similar solutions
and intermediate asymptotics has long been recognized \cite{Barenblatt}.
Moreover, traveling waves have previously been identified as fixed points of
suitably defined RG transformations \cite{paquette1994structural,
chen1995numerical}. The present work therefore examines whether such
traveling-wave fixed points are necessarily associated with universality
classes through the same irrelevant-structure elimination mechanism that
underlies universality in ordinary self-similar fixed points.

In the present work, we address this question by extending the predictive RG
framework to traveling-wave solutions. As representative examples, we
consider Burgers equation and the Korteweg--de Vries (KdV) equation. These
classical traveling-wave systems provide complementary settings in which to
test whether the universality-class formation mechanism identified
previously for self-similar solutions also operates for traveling-wave fixed
points. We find instead that the logarithmic scaling structure underlying
traveling-wave fixed points prevents the emergence of the irrelevant
structures whose elimination normally generates universality classes. This
leads to the central conclusion of the paper: RG fixed points need not be
associated with universality classes. Within the same RG framework, the
conventional construction of traveling waves is recast as a fixed-point
problem for the RG flow, while the logarithmic variable used to map
traveling waves onto self-similar forms acquires a natural interpretation as
the time-like parameter of the continuous RG flow, up to an additive
constant.

The absence of universality classes has an important consequence. If
irrelevant structures are not eliminated, information that would normally be
discarded under RG transformations may instead persist asymptotically. This
observation suggests a natural link between universality-class formation and
memory retention, which will be explored through the Burgers and KdV
examples.

\begin{figure}[tbp]
\includegraphics[width=\columnwidth]{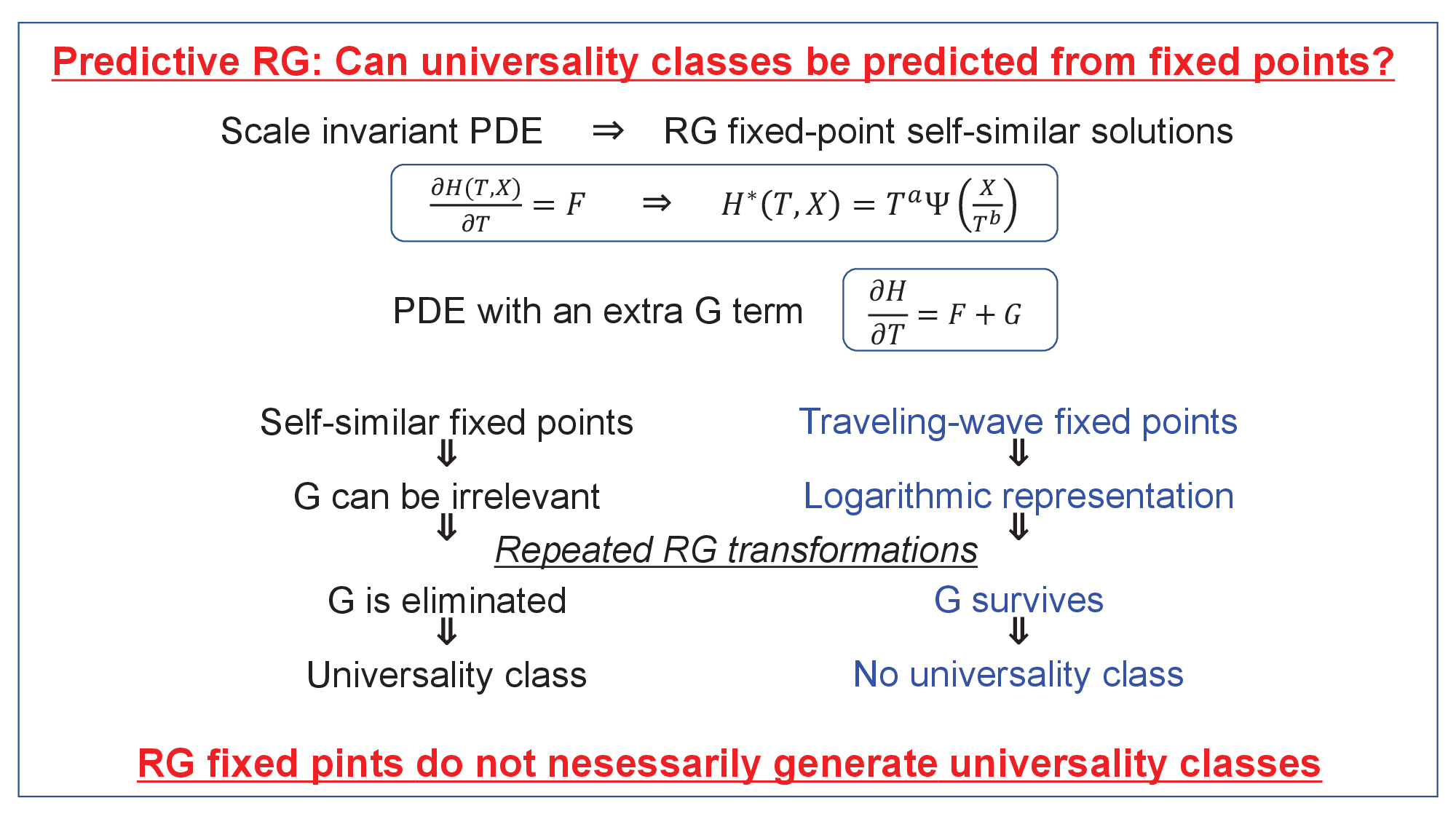}
\caption{Conceptual relation between universality-class formation in the
predictive RG framework and the traveling-wave fixed points considered in
the present work. For self-similar fixed points, repeated RG transformations
may eliminate irrelevant structures, leading to universality classes. By
contrast, for traveling-wave fixed points the logarithmic representation
prevents the emergence of irrelevant structures, so that repeated RG
transformations do not generate universality classes. The figure summarizes
the central conclusion of the present work: RG fixed points do not
necessarily generate universality classes.}
\label{Fig1}
\end{figure}

The central conceptual relation between universality-class formation in the
predictive RG framework and the traveling-wave fixed points considered in
the present work is summarized schematically in Fig \ref{Fig1}, emphasizing
the contrast between irrelevant-structure elimination and its absence under
repeated RG transformations.

\section{Unified RG framework}

The unified RG framework \cite{Okumura2025RG, okumura2026combined,
Okumura2026nonlinear} has been developed for a generic PDE of the form $%
\partial _{T}\mathbf{H}(T;\mathbf{X})$ $=$ $\mathbf{F}(\mathbf{H},D_{1}%
\mathbf{H},D_{2}\mathbf{H},\cdots )$ $+$ $\mathbf{G}(\mathbf{H},D_{1}\mathbf{%
H},D_{2}\mathbf{H},\cdots )$ where $D_{i}\mathbf{H}$ stands for the $i$-th
spatial derivative (e.g., $D_{2}\mathbf{H}$ stands for the set $\{\partial
_{X_{i}}\partial _{X_{j}}H_{k}\}$). In the following, for the present
analysis, we briefly summarize the framework in a simple case in which the
vectors $\mathbf{H}$, $\mathbf{G}$, and $\mathbf{X}$ all have only a single
component and thus we set $\mathbf{H}\rightarrow H$, $\mathbf{G}\rightarrow
G $, and $\mathbf{X}\rightarrow X$ in the PDE: 
\begin{equation}
\partial _{T}H(T;X)=F(H,D_{1}H,D_{2}H,\cdots )+G(H,D_{1}H,D_{2}H,\cdots ).
\label{eq13}
\end{equation}

In such a simple case, the framework introduces a scale transformation for
the scale factor $L$ and the exponent $B$, which are both positive:

\begin{eqnarray}
T^{\prime } &=&T/L^{B},\text{ }X^{\prime }=X/L \\
H^{\prime }(T^{\prime };X^{\prime }) &=&L^{A}H(T,X)\text{ }\equiv
H_{L}(T^{\prime },X^{\prime })  \label{eq15} \\
&\Leftrightarrow &\text{ }H_{L}(T,X)=L^{A}H(L^{B}T,LX).  \label{eq15B}
\end{eqnarray}%
\newline
The scale factor $L$ satisfies $L>1$ for \textit{long-time asymptotics} as
in the present case where \textit{large scale physics} is important (Case II
in the previous study \cite{Okumura2026nonlinear}). The terms in $F$ and $G$
are regarded as \textit{scale-invariant}, \textit{relevant}, and \textit{%
irrelevant} depending on whether the exponent $M$ defined in Eq \ref{M1} in
Appendix \ref{A-U} is zero, positive, and negative, respectively, for Case
II.

If we consider the case in which the terms in $F$ are all scale invariant
while the terms in $G$ are non-scale-invariant, the latter terms are
eliminated as the RG transformation defined in Eq \ref{eq17} is repeatedly
applied, and thus called \textit{irrelevant }(the former scale-invariant
terms are \textit{not} eliminated): When the irrelevant terms in $G$ are
eliminated under repeated RG transformations, different PDEs can flow toward
the same scale-invariant PDE and share the same self-similar RG fixed point.
These PDEs then constitute a universality class \cite{Okumura2026nonlinear}.

As detailed in Appendix \ref{A-U}, the framework shows that \textit{%
self-similar solutions emerge as RG fixed points if }$h^{\ast }(X)$ \textit{%
exists}: 
\begin{equation}
H^{\ast }(T;X)=T^{-A/B}h^{\ast }(\xi )\text{ with }\xi =X/T^{1/B}
\label{e20}
\end{equation}%
where $h^{\ast }(X)$ can be obtained as a stationary solution to the RG flow
equation given in Eq \ref{eq21}, i.e., the fixed-point equation: 
\begin{equation}
0=Ah^{\ast }(X)+BF(h^{\ast },D_{1}h^{\ast },D_{2}h^{\ast },\cdots )+X\frac{%
\partial h^{\ast }(X)}{\partial X},  \label{eq21A2}
\end{equation}%
It is stressed here that, as described in Appendix \ref{A-U}, the scale
invariance of $F$ is required for fixed points, i.e., self-similar solutions
to exist.

\section{\noindent Burgers shock waves}

We consider the long time asymptotics of Burgers shock wave equation:%
\begin{equation}
\partial _{t}V(t,x)+V\partial _{x}V=\nu \partial _{x}^{2}V  \label{eq01}
\end{equation}%
with an initial front localized in the vicinity $x=x_{f}$: near this
position $V(0,x)$ takes a value between $a$ and $b$ ($a>b>0$) such that 
\begin{equation}
V(0,x)\rightarrow \left\{ 
\begin{array}{ccc}
b &  & x\rightarrow \infty \\ 
a &  & x\rightarrow -\infty%
\end{array}%
\right.  \label{eq01b}
\end{equation}%
We seek a traveling-wave solution of the form, i.e., the traveling-wave
ansatz, 
\begin{equation}
V(t,x)=\widehat{V}(\zeta =x-vt+k)  \label{eq01c}
\end{equation}%
with the front moving with a velocity $v$.

By introducing logarithmic variables:%
\begin{equation}
t=\log T-\log T_{0}\text{; }x=\log X  \label{eq02}
\end{equation}%
we obtain, with $T_{0}^{-v}=K$, 
\begin{equation}
x-vt+k=\log (\frac{X}{KT^{v}}).
\end{equation}%
In terms of new variables $T$ and $X$ with $V(t,x)=H(T,X)$, \textit{the
traveling-wave solution can be regarded as a self-similar solution}, $H(T,X)=%
\widehat{H}(\frac{X}{KT^{v}})$. Comparison with the generic RG fixed-point
form in Eq \ref{e20} already suggests that $A=0$, since the power-law
prefactor $T^{-A/B}$ is absent. The scaling analysis below confirms this
expectation.

In terms of $T$ and $X$, Burgers equation takes the form%
\begin{equation}
T\partial _{T}H=F[H]  \label{eq03}
\end{equation}%
with

\begin{equation}
F[H]=-XH\partial _{X}H+\nu (X\partial _{X}H+X^{2}\partial _{X}^{2}H).
\label{eq04}
\end{equation}

By making the replacements, $T\rightarrow L^{B}T$ and $X\rightarrow LX$,
with $H_{L}(T,X)=L^{A}H(L^{B}T,LX)$, for example, the term $XH\partial _{X}H$
is transformed as: $XH\partial _{X}H$ $\rightarrow $ $LXH(L^{B}T,LX)L^{-1}%
\partial _{X}H(L^{B}T,LX)$ $\rightarrow $ $L^{-2A}XH_{L}\partial _{X}H_{L}$,
where we have used Eq \ref{eq15B} in the second transformation. Likewise, we
get $T\partial _{T}H\rightarrow L^{-A}T\partial _{T}H_{L}$ and 
\begin{equation}
X^{n}\partial _{X}^{n}H\rightarrow L^{-A}X^{n}\partial _{X}^{n}H_{L}\text{
with }n=1,2,\ldots .  \label{eqA}
\end{equation}%
From these transformation rules, requiring the scale invariance of Eq \ref%
{eq04} results in $A=0$, in agreement with the absence of the prefactor $%
T^{-A/B}$ anticipated from the fixed-point form in Eq \ref{e20}. In other
words, the existence condition of fixed points makes $A=0$.

To investigate the long-time behavior, we apply the RG transformation for $%
L>1$ repeatedly. The unified RG framework gives the RG-fixed point
self-similar solution as%
\begin{equation}
H^{\ast }(T,X)=h^{\ast }(\xi )\text{ with }\xi =\frac{X}{KT^{1/B}}
\label{eq05}
\end{equation}%
To make the connection between repeated RG transformations and a continuous
RG flow explicit, we introduce the logarithmic parameter $\tau =B\log L$.
Repeated RG transformations then correspond to successive increments in $%
\tau $, which becomes a continuous RG-flow parameter in the
infinitesimal-step limit $L\rightarrow 1$, as summarized in Appendix \ref%
{A-U}. We then obtain $B\frac{dH_{L}(T,X)}{d\tau }$ $=AH_{L}$ $+X\frac{%
\partial H_{L}}{\partial X}$ $+BF[H_{L}]$, in which we set $T=1$ and $%
H_{L}(1,X)=R_{L}f(X)\equiv \widehat{h}(\tau ,X)$, to get the RG flow
equation: 
\begin{equation}
B\frac{d\widehat{h}(\tau ,X)}{d\tau }=X\frac{\partial \widehat{h}}{\partial X%
}+BF[\widehat{h}],  \label{eq07}
\end{equation}%
which corresponds to Eq \ref{eq21}. The stationary solution $h^{\ast }(\tau
,X)=f(X)$ to the flow equation when $\frac{d\widehat{h}(\tau ,X)}{d\tau }=0$
satisfies the nonlinear second-order ordinary differential equation (ODE),
i.e., the fixed-point equation for the Burgers equation: 
\begin{equation}
\frac{X}{B}\partial _{X}f=Xf\partial _{X}f-\nu (X\partial
_{X}f+X^{2}\partial _{X}^{2}f).  \label{eq08}
\end{equation}%
Notably, under the spatial logarithmic transformation in Eq \ref{eq02}, this
Burgers fixed-point equation is transformed into an ODE obtained from the
governing equation with the conventional traveling-wave ansatz (see Appendix %
\ref{A-2}). The RG formulation therefore recasts the standard traveling-wave
construction as a fixed-point problem for the RG flow.

We seek the solution to Eq \ref{eq08} of the form 
\begin{equation}
f(X)=\frac{a^{\prime }+b^{\prime }X^{c}}{1+X^{c}},  \label{A1}
\end{equation}%
where $a^{\prime }$, $b^{\prime }$, and $c$ are constants associated with
the fixed-point family. Note that this form is consistent with a front-type
solution, satisfying the remote conditions $f\rightarrow a^{\prime }$ and $%
b^{\prime }$ as $X\rightarrow 0$ and $\infty $, respectively.

Substitution into Eq \ref{eq08} yields $\frac{1}{B}=a^{\prime }-c\nu
=b^{\prime }+c\nu $. When the matching conditions with the remote states
defined in Eq \ref{eq01b} with $h^{\ast }=f$, the constants $a^{\prime }$
and $b^{\prime }$ are identified $a$ and $b$ in Eq \ref{eq01b}. This leads to%
\begin{equation}
c=\frac{a-b}{2\nu }\equiv \frac{1}{l}\text{ and }\frac{1}{B}=\frac{a+b}{2}%
\equiv v.  \label{eq10}
\end{equation}
Thus, if $a\neq b$, we obtain, from Eq \ref{eq05}, $H^{\ast }(T,X)=\frac{%
a+b\xi ^{1/l}}{1+\xi ^{1/l}}$ with $\xi =\frac{X}{KT^{v}}$, i.e.,%
\begin{equation}
V(t,x)=\frac{a+be^{\frac{x-vt+k}{l}}}{1+e^{\frac{x-vt+k}{l}}}  \label{eq12}
\end{equation}%
The solution in Eq \ref{eq12} satisfies the same remote conditions as the
initial condition in Eq \ref{eq01b}, and describes a front of width $l$,
which moves with the velocity $v$. This result reproduces the classical
viscous Burgers shock solution \cite{Barenblatt} (see Appendix \ref{A-2}),
thereby providing a nontrivial validation of the unified RG formulation.

Interestingly, the solution in Eq \ref{eq12} retains memory of the remote
asymptotic states $a$ and $b$, as well as the system parameter $\nu $. More
precisely, the velocity $v$ is determined solely by the remote states $a$
and $b$, whereas the width $l$ additionally depends on $\nu $. Thus,
although the asymptotic shock retains information about the initial
condition through its remote states, the detailed shape of the initial front
does not enter the selected traveling-wave parameters. In this sense, the
Burgers dynamics retains only a reduced form of memory of the initial
condition, encoded through the remote matching data.

As seen above, the exponent $A$ is determined by a dimensional analysis,
while the exponent $B$, or the velocity $v$, is determined when we seek the
stationary solution of the RG flow equation, with using remote matching
conditions, as in the conventional construction (see Appendix \ref{A-2}).

More importantly, the RG analysis reveals that $A=0$, which makes any
analytic term $G$ added to the governing equation scale invariant, as shown
below. This can be seen directly by considering a generalized equation $%
\partial _{t}V$ $+V\partial _{x}V$ $=\nu \partial _{x}^{2}V$ $+G$. Here, $G$
is assumed to be analytic in $V$ and its spatial derivatives, by which we
mean that $G$ can be expanded as a linear combination of terms of the form 
\begin{equation}
V^{N_{0}}(\partial _{x}V)^{N_{1}}(\partial _{x}^{2}V)^{N_{2}}\cdots
(\partial _{x}^{m}V)^{N_{m}}.  \label{eq16}
\end{equation}%
Strictly speaking, we consider only the regular-analytic case, which means $%
m $ is finite. Under the change of variables in Eq \ref{eq02}, for example,
the term with $m=3$ is transformed into $H^{N_{0}}$ $(X\partial
_{X}H)^{N_{1}}$ $(X\partial _{X}H+X^{2}\partial _{X}^{2}H)^{N_{2}}(X\partial
_{X}H+3X^{2}\partial _{X}^{2}H+X^{3}\partial _{X}^{3}H)^{N_{3}}$. In
general, we see that the logarithmic coordinate transformation converts
every local derivative of order $m$ into a linear combination of operators $%
X^{k}\partial _{X}^{k}${} ($k=1,2,..,m$), whose scaling dimension exactly
compensates the derivative dimension. Consequently, from Eq \ref{eqA}, the
term in Eq \ref{eq16} is transformed into the term of the form, $L^{M_{0}}$ $%
H_{L}^{N_{0}}$ $(X\partial _{X}H_{L})^{N_{1}}$ $(X\partial
_{X}H_{L}+X^{2}\partial _{X}^{2}H_{L})^{N_{2}}\cdots $, with $%
M_{0}=-A(N_{0}+N_{1}+\cdots +N_{m})$. Therefore, when $A=0$, every term in
this analytic expansion is scale invariant.

Hence $G$ remains scale invariant rather than being progressively eliminated
as an irrelevant structure under repeated RG transformations. Consequently,
the usual RG mechanism responsible for universality classes is absent.\ This
conclusion applies to the class of analytic local terms defined above; more
general nonlocal or nonanalytic terms are outside the scope of the present
analysis.

Notably, the scale invariance discussed above holds for the finite RG
transformation characterized by the finite factor $L$. Thus, the scale
invariance in the present context should be distinguished from that
marginality inferred from a linearized RG analysis often discussed in
critical phenomena.

\section{KdV solitary wave}

Consider the KdV equation:\textit{\ }%
\begin{equation}
\partial _{t}V+V\partial _{x}V+\beta \partial _{x}^{3}V=0  \label{eq1}
\end{equation}%
By introducing variables as in Eq \ref{eq02}, the KdV equation changes into
Eq \ref{eq03} with

\begin{equation}
-F[H]=XH\partial _{X}H+\beta (X^{3}\partial _{X}^{3}H+3X^{2}\partial
_{X}^{2}H+X\partial _{X}H)  \label{eq20}
\end{equation}%
As before, by making the replacements, $T\rightarrow L^{B}T$ and $%
X\rightarrow LX$, with $H_{L}(T,X)=L^{A}H(L^{B}T,LX)$, we require the scale
invariance of Eq \ref{eq03} with Eq \ref{eq20}, to conclude $A=0$. 

As in the Burgers case, the resulting vanishing scaling dimension makes
analytic term $G$ added to the right-hand side of Eq \ref{eq1} scale
invariant, so that the usual irrelevant-structure mechanism for generating
universality classes is absent. We therefore expect the same absence of
universality-class formation to hold for a broad class of traveling-wave
equations containing analytic local terms of this type.

Then, we obtain the RG fixed point as in Eq \ref{eq05} and the RG flow
equation as in Eq \ref{eq07}. The stationary solution $h^{\ast }(\tau
,X)=f(X)$ for the latter satisfies the following fixed-point equation for
the KdV equation:%
\begin{equation}
\frac{X}{B}\partial _{X}f=Xf\partial _{X}f+\beta (X^{3}\partial
_{X}^{3}f+3X^{2}\partial _{X}^{2}f+X\partial _{X}f)  \label{eq24}
\end{equation}%
As seen in the Burgers case, under the spatial logarithmic transformation in
Eq \ref{eq02}, this KdV fixed-point equation is transformed into an ODE
obtained from the governing equation and the traveling-wave ansatz (see
Appendix \ref{A-2 copy(1)}). We again observe that the RG formulation
recasts the standard traveling-wave construction as a fixed-point problem
for the RG flow.

If we seek a solution of the form 
\begin{equation}
f(X)=\frac{12v}{(X^{C}+X^{-C})^{2}},  \label{A3}
\end{equation}%
which belongs to a continuous fixed-point family parameterized by $C$. Note
that this form is consistent with a soliton-type solution, satisfying the
remote conditions $f\rightarrow 0$ as $X\rightarrow \pm \infty $.

Substituting the form into Eq \ref{eq24}, we find%
\begin{equation}
\frac{1}{B}=v=4\beta C^{2}\text{ }  \label{eq26}
\end{equation}%
and obtain, from Eq \ref{eq05}, $H^{\ast }(T,X)=\frac{12v}{(\xi ^{\sqrt{%
\frac{v}{\beta }}/2}+\xi ^{-\sqrt{\frac{v}{\beta }}/2})^{2}}$ with $\xi =%
\frac{X}{KT^{v}}$. Since $C$ is an arbitrary positive constant, Eq \ref{eq24}
admits a continuous family of fixed-point solutions parameterized either by $%
C$ or, equivalently, by the velocity $v$.

In contrast to the Burgers case, the RG fixed-point equation does not
determine a unique velocity. We note that this solution $H^{\ast }(T,X)$ is
small if $\xi $ is either large or small, representing a solitary wave (%
\textit{soliton}) that peaks at $X=KT^{v}$. In this manner, we obtain%
\begin{equation}
V(t,x)=\frac{3v}{\cosh ^{2}(\frac{x-vt+k}{2l})}  \label{eq28}
\end{equation}%
with the length $l=\sqrt{\frac{\beta }{v}}$, which reproduces a known result
as discussed below, thereby providing another nontrivial validation of the
unified RG formulation.

A deeper understanding of this problem was provided by Gardner, Greene,
Kruskal and Miura \cite{gardner1967method, gardner1974korteweg}. For a given
initial distribution $V(0,x)$ that decays rapidly as $x\rightarrow \pm
\infty $, the initial condition $V(0,x)$ and the system parameter $\beta $
determine an associated Schr\"{o}dinger problem. The number $N$ of its
negative discrete eigenvalues gives the number of solitons appearing in the
long-time asymptotics, while the eigenvalues determine their velocities $%
v=v_{n}$ ($n=1,2,\cdots ,N$){}. The initial-value problem also fixes the
corresponding position parameters $k=k_{n}${}.

The RG fixed-point equation therefore yields a continuous family of
admissible soliton fixed points in Eq \ref{eq28} with $v>0$. The initial
condition and system parameter determine which member (or members) of this
family are realized in the long-time asymptotics.

As in the Burgers case, the RG formulation thus recasts the conventional
traveling-wave construction with the traveling-wave ansatz as a fixed-point
problem for the RG flow. In contrast to the Burgers case, however, the
remote condition does not select the velocity of a KdV soliton (see Appendix %
\ref{A-2 copy(1)}). Instead, the realized velocities are determined by the
spectrum associated with the initial condition. Consequently, both the
soliton velocity $v$ and width $l$=$\sqrt{\beta /v}${} retain explicit
dependence on the initial state $V(0,x)$ and the system parameter $\beta $.
The KdV soliton therefore provides a clear example of an RG fixed point
without a universality class, with characteristic parameters of the
asymptotic state retaining memory of both the initial condition and the
system parameter. This memory is qualitatively stronger than in the Burgers
case: whereas the Burgers shock parameters are fixed by the remote
asymptotic states, the KdV soliton velocities are selected through spectral
data associated with the full initial profile $V(0,x)$. Thus, details of the
initial condition that are asymptotically forgotten in the Burgers shock
problem can remain encoded in the KdV soliton content.\newline

\section{Discussion}

The RG formulation recasts the conventional construction based on the
governing equation and a traveling-wave ansatz as a fixed-point problem for
the RG flow, thereby allowing traveling waves to be examined within the same
framework previously used to identify universality classes for self-similar
solutions. For the Burgers and KdV equations, the classical shock and
soliton solutions emerge as stationary RG fixed points, with the parameter
structure allowed by the conventional formulation: the remote conditions fix
the Burgers shock velocity and width, whereas in the KdV case they specify
only a family of admissible soliton solutions. Thus, the RG fixed-point
formulation provides an alternative derivation of the classical
traveling-wave solutions based on the traveling-wave ansatz.

More importantly, the present results reveal a fundamental distinction
between RG fixed points and universality classes. For the traveling-wave
systems considered here, the logarithmic representation places the field and
all of its derivatives on the same scaling footing. Because scale invariance
then requires this common scaling dimension to vanish ($A=0$), any analytic
term $G$ added to the governing equation remains scale invariant rather than
becoming irrelevant under repeated RG transformations. Within the unified
predictive RG framework, universality classes emerge through the elimination
of irrelevant structures. That mechanism is therefore absent in the
traveling-wave systems considered here, even though RG fixed points continue
to exist. More broadly, this suggests that the association between RG fixed
points and universality classes, often taken for granted in RG analyses, can
break down for a broad class of traveling-wave systems with the scaling
structure considered here. From this perspective, the absence of
universality classes is not a technical consequence of the logarithmic
transformation but reflects the underlying traveling-wave scaling structure
made explicit by the logarithmic representation. The distinct question of
whether RG fixed-point stability determines front selection is considered
separately in Ref \cite{Okumura2026FKPP}.

The Burgers and KdV examples further suggest a possible connection between
memory retention and the absence of the usual universality-generating
mechanism. The absence of universality classes implies that system-specific
information is not fully washed out asymptotically. It is therefore natural
to expect a greater tendency for memory retention, although the degree and
form of the retained information remain system dependent. This tendency is
particularly evident in the KdV case, where information associated with the
initial condition survives in the spectral data that determine the soliton
parameters. By contrast, the Burgers shock retains only the information
contained in the remote asymptotic states, so that details of the initial
profile are lost. Thus, the two examples exhibit different degrees and forms
of memory despite sharing the absence of the usual universality-generating
mechanism.

This form of memory retention differs from that encountered in
memory-retaining self-similar fixed points \cite{Okumura2026memory}, where
additional information remains explicitly within the fixed-point
representation itself, for example through fixed-point forms such as $\eta
^{\alpha }f^{\ast }(\xi /\eta ^{\beta })$, where $\xi $ is the similarity
variable and $\eta $ is an anomalous-scaling variable that retains memory of
the initial conditions. By contrast, in the traveling-wave examples
considered here, the retained information does not appear as anomalous
scaling variables within the fixed-point representation. Instead, it
determines which member of a fixed-point family is realized, through
remote-state matching in the Burgers case and through spectral data in the
KdV case. The retained information therefore enters through
fixed-point-family selection rather than through anomalous scaling variables
in the fixed-point representation itself.

Interestingly, the logarithmic transformation that maps traveling waves onto
self-similar solutions also provides the natural time-like parameter for the
continuous RG flow. Indeed, under the RG scale relation $T=L^{B}$, the
definition $\tau =B\log L$, together with $t=\log T-\log T_{0}$ in Eq (\ref%
{eq02}), gives $\tau =t+\log T_{0}$ (see Appendix \ref{A-U}). Thus, apart
from a constant shift of the origin, the logarithmic time $t$ used to
represent a traveling wave in self-similar form coincides with the time-like
parameter $\tau $ of the RG flow. In other words, the conventional
logarithmic transformation used to map traveling waves onto self-similar
solutions acquires a direct interpretation as the transformation to RG-flow
time.

\section{\noindent \noindent Conclusion\label{Conclusion}}

We have shown that the conventional traveling-wave construction based on
matching can be recast within a predictive RG framework as an RG fixed-point
problem, reproducing the classical Burgers shock and KdV soliton solutions.
In this formulation, the conventional logarithmic transformation that maps
traveling waves onto self-similar solutions acquires a direct interpretation
as the transformation to RG-flow time. The Burgers and KdV cases further
reveal different degrees of memory retention, ranging from dependence on
remote asymptotic states to more detailed dependence on the initial profile
through spectral data. This contrast suggests a possible connection between
memory retention and the absence of the usual universality-generating
mechanism, although the latter does not determine how much memory is
retained.

More fundamentally, the present analysis shows that traveling-wave RG fixed
points need not generate universality classes. Although such fixed points
exist and reproduce the classical Burgers and KdV traveling waves, the usual
irrelevant-structure elimination mechanism is absent. The Burgers and KdV
examples suggest that the absence of universality classes is not merely a
peculiarity of particular equations but may reflect a generic feature of
traveling-wave RG fixed points formulated through logarithmic coordinates.
This identifies a fundamental limitation of universality-class formation
within the predictive RG framework. \newline
\textit{The author is grateful to Professor Nigel Goldenfeld (UCSD) for
helpful comments and encouragement. This work was supported by JSPS\ KAKENHI
Grant Number JP24K00596. }\newline
\textit{Declaration of generative AI and AI-assisted technologies in the
manuscript preparation process: During the preparation of this work, the
author used Microsoft Copilot to assist with language editing, organization,
and clarity of presentation. The author reviewed and edited all AI-assisted
content as needed and takes full responsibility for the content of the
publication.}\newline
\textit{Data Availability Statement: This study is theoretical in nature and
did not generate or analyze any datasets.}

\appendix
\clearpage

\section{Supplementary notes\label{A2}}

\subsection{Note on the unified RG framework\label{A-U}}

The detailed derivation of the unified RG framework is given in the previous
article \cite{Okumura2026nonlinear}. For the purposes of the present work,
we summarize here the necessary information. Under the scale transformation
given in Eq \ref{eq15}, a term in $F$ or $G$ in Eq \ref{eq13} of the form, $%
H^{N_{0}}(\partial _{X}H)^{N_{1}}(\partial _{X}^{2}H)^{N_{2}}\cdots $, is
changed into the form, $L^{M}H_{L}^{N_{0}}(\partial
_{X}H_{L})^{N_{1}}(\partial _{X}^{2}H_{L})^{N_{2}}\cdots $, with the \textit{%
scaling factor} $L^{M}$ characterized by the \textit{scaling exponent} $M$,
which defines relevance of the term:%
\begin{equation}
M=A+B-[N_{0}A+N_{1}(A+1)+N_{2}(A+2)+\cdots ],  \label{M1}
\end{equation}

In the present case, we are interested in \textit{intermediate asymptotics}
(Case II in the previous article \cite{Okumura2026nonlinear}) with the
boundary condition at $T=1$: $H(1,X)=h(X)$. The RG transformation is defined
for $h(X)$ as%
\begin{equation}
\emph{R}_{L,G}h(X)\equiv L^{A}H(L^{B};LX)=H_{L}(1;X)\text{.}  \label{eq17}
\end{equation}%
The second equality is based on Eq \ref{eq15B}.

If $F$ is scale-invariant and $G$ is irrelevant, we can expect that repeated
RG transformations make $h(X)$ and Eq \ref{eq13} flow toward their fixed
points: $h^{\ast }(X)$ and the corresponding scale-invariant equation with $%
G=0$, where the fixed point is defined by the following equation: $\emph{R}%
_{L,G^{\ast }}h^{\ast }(X)=h^{\ast }(X)\Leftrightarrow L^{A}H^{\ast
}(L^{B};LX)=h^{\ast }(X)$. If such a point exists, setting $T=L^{B}$ in this
equation results in $T^{A/B}H(T;T^{1/B}X)=h^{\ast }(X)$, from which we
obtain Eq \ref{e20}: $H^{\ast }(T;X)=T^{-A/B}h^{\ast }(X/T^{1/B})$.

To connect repeated discrete RG transformations with a continuous RG flow,
we introduce the logarithmic variable $\tau =B\log L$ for Case II. For $G=0$%
, after $n$ RG steps, the corresponding scale factor $L^{n}$ is represented
by $\tau _{n}=B\log L^{n}=n\tau $, so that the logarithmic variable converts
the multiplicative sequence of RG scales into an additive discrete
parameter. Taking the infinitesimal-step limit $L\rightarrow 1$, then yields
the continuous RG flow equation for $\widehat{h}(\tau ;X)\equiv H_{L}(1;X)=%
\emph{R}_{L}f(X)$:

\begin{equation}
B\frac{d\widehat{h}(\tau ;X)}{d\tau }=A\widehat{h}(\tau ;X)+BF(\widehat{h}%
,D_{1}\widehat{h},\cdots )+X\frac{\partial \widehat{h}(\tau ;X)}{\partial X}.
\label{eq21}
\end{equation}%
By setting $\frac{d\widehat{h}(\tau ;X)}{d\tau }=0$ in Eq \ref{eq21}, we
obtain Eq \ref{eq21A2} for the stationary solution $h^{\ast }(X)$. Its
stability can be examined by substituting $\widehat{h}(\tau ;X)=h^{\ast
}(X)+\delta (X)e^{\omega \tau }$ into Eq \ref{eq21}, linearizing the
equation in terms of $\delta (X)$, and examining the sign of $\omega $.
Since $\tau $ tends to $+\infty $ under repeated RG transformations, a
negative (positive) $\omega $ corresponds to a \textit{stable} (an \textit{%
unstable}) mode.

\subsection{Note on Burgers shock wave\label{A-2}}

Substituting the traveling-wave ansatz for $V$, Eq \ref{eq01c}, into Eq \ref%
{eq01}, we have the following ODE:%
\begin{equation}
-v\widehat{V}^{\prime }(\zeta )+\widehat{V}(\zeta )\widehat{V}^{\prime
}(\zeta )=\nu \widehat{V}^{\prime \prime }(\zeta ).  \label{eq14}
\end{equation}%
Note that this ODE is equivalent to the RG flow equation given in Eq \ref%
{eq08}, under the logarithmic transformation.

Integrating Eq \ref{eq14} once with the matching condition $\widehat{V}=b$
as $\zeta \rightarrow \infty $ as in Eq \ref{eq01b}, we obtain $\nu \frac{d%
\widehat{V}}{d\zeta }=-v(\widehat{V}-b)+\frac{\widehat{V}^{2}-b^{2}}{2}$,
and matching $\widehat{V}$ at $\zeta \rightarrow -\infty $ then gives $%
-v(a-b)+\frac{a^{2}-b^{2}}{2}=0$, i.e., $v=\frac{a+b}{2}$ as before. From
this, we obtain $\nu \frac{d\widehat{V}}{d\zeta }=-\frac{(a-\widehat{V})(%
\widehat{V}-b)}{2}$. Notably, this ordinary differential equation can be
directly solved by separation of variables to recover Eq \ref{eq12}. As seen
above, the matching conditions identify the fixed-point parameters with the
remote asymptotic states of the traveling-wave solution.

\subsection{Note on KdV equation\label{A-2 copy(1)}}

Substituting the traveling-wave ansatz for $V$, Eq \ref{eq01c}, into Eq \ref%
{eq1}, we have the following ODE:

\begin{equation}
-v\widehat{V}^{\prime }(\zeta )+\widehat{V}(\zeta )\widehat{V}^{\prime
}(\zeta )+\beta \widehat{V}^{^{\prime \prime \prime }}(\zeta )=0.
\label{eq29}
\end{equation}%
As in the Burgers case, this ODE is equivalent to the RG flow equation given
in Eq \ref{eq24}, under the logarithmic transformation.

Integrating Eq \ref{eq29} once using the boundary condition $\widehat{V}=0$
as $\zeta \rightarrow \infty $, we obtain $\beta \frac{d^{2}\widehat{V}}{%
d\zeta ^{2}}-v\widehat{V}+\frac{\widehat{V}^{2}}{2}=0$. Setting $\widehat{V}%
=1/U^{2}$, we get $\beta (6U^{\prime 2}-2UU^{^{\prime \prime }})-vU^{2}+%
\frac{1}{2}=0$ and seek the solution of the form $U=A\cosh (B\zeta )$ to
find $A^{2}=\frac{1}{3v}$ and $B^{2}=\frac{v}{4\beta }$, to recover Eq \ref%
{eq28}.

As seen above, the asymptotic condition determines the localized soliton
family but does not determine the soliton velocity. The latter is selected
through the spectral data of the initial-value problem.

\end{document}